# Magnetization near a constriction between BCS superconductors by spin-dependent tunneling

O. Entin-Wohlman,[1] R. I. Shekhter,[2] M. Jonson,[2] and A. Aharony[1]

[1]*School of Physics and Astronomy, Tel Aviv University, Tel Aviv 6997801, Israel*[*]
[2]*Department of Physics, University of Gothenburg, SE-412 96 Göteborg, Sweden*
(Dated: November 4, 2023)

Spin-dependent electron tunneling through a voltage-biased micro-constriction between two bulk superconductors is shown to create a dc component of the magnetization in the superconductors near the constriction and an ac Josephson-like spin current. The static magnetization appears in one superconductor even if the other is replaced by a normal conductor. Although spin-dependent tunneling generates quantum spin fluctuations also in the absence of a bias, the formation of spin-triplet Cooper pairs, necessary for the creation of magnetization, is blocked by destructive interference between different quasi-electron and quasi-hole tunneling channels, unless there is an asymmetry between the tunneling densities of states for electrons and holes. Breaking the symmetry in the electron-hole tunnel density of states and creating electron-hole tunneling imbalance by biasing the device destroys the destructive interference and enables triplet Cooper-pair formation. As a result, magnetizing the superconductor becomes possible. The role of the voltage in lifting the blockade hindering the spin-triplet Cooper pair formation is an example of an electrically controlled dissipation-less spintronic phenomenon.

*Introduction* — Tunable magnetizations play an obvious role in spintronics and in quantum information processing, which is why the question addressed in this work, how to magnetize a superconductor, is of great current interest. In its ground state, the BCS wave function of a homogeneous superconductor describes Cooper-paired electron states. These states are time reversed with respect to each other and are eigenstates of an operator that projects the electronic spin on a certain axis, denoted here as the $z$-axis, which is defined by the order parameter (pairing potential) that induces the pairing. The nature of this superconducting pairing can change significantly in spin-inhomogeneous materials such as structures with paramagnetic impurities, conductors affected by spatially inhomogeneous magnetic fields, hybrid superconducting structures comprising ferromagnetic parts, *etc.* Examples are the suppression of superconductivity in alloys with paramagnetic impurities [1], the triplet proximity effect at superconductor-ferromagnet interfaces or spin-orbit active normal-superconducting interfaces [2–6], and anomalous Josephson effects in hybrid superconducting structures [7–10]. Many such interfaces are listed inn Eschrig's review [11].

Two bulk superconductors connected by a weak link in the form of a micro-constriction [Fig. 1] is an example of a hybrid structure, where an inhomogeneity affects the superconducting properties locally on the scale of the diameter $a$ of the constriction cross-section. The extent of the influence of the constriction depends on the relation between $a$ and the coherence length $\xi$, which is the length scale of the non-locality of superconducting correlations. This is because the superconducting order-parameter at any point in space is determined self-consistently in a way that involves all electrons within the coherence length $\xi$ around the point. If $\xi \gg a$ the order parameter close to the constriction is mostly determined by bulk electron pairs located far away, in a region where there are very few electrons that have tunneled through the constriction. This is due to the fact that the density of tunneled electrons diminishes with the distance $r$ from the point contact as $1/r^2$ (for a three-dimensional superconductor). In this case we are allowed to use the bulk value $\Delta$ of the homogeneous superconductor's order parameter even in the vicinity of the weak link and assume that the order parameter pairs electrons that are eigenstates of the spin operator $\hat{s}_z$.

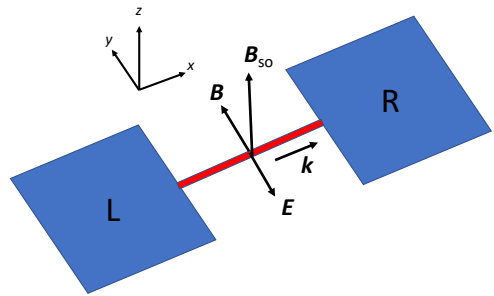

FIG. 1. (color online) Sketch of the system considered. A spin-orbit active one-dimensional weak link of length $d$(red) bridges two superconducting leads, L and R (blue). The tunneling amplitude $\mathcal{J}$ given by Eq. (11) is obtained if an external electric field, applied in the negative $y$-direction, interacts with an electron moving along the $x$-direction with momentum $\mathbf{k}$ and generates a pseudo-magnetic field $\mathbf{B}_{so}$ in the $z$-direction. An external magnetic field $\mathbf{B}$ should also be applied in the $y$-direction.

Depending on the nature of the weak link, we may have different scenarios for electron transfer between the superconductors through the constriction. Here we consider the mechanism of spin-dependent tunneling. The spin dependence may arise from, e.g., scattering off magnetic inclusions or spin-orbit interactions in the weak link (see below). Our results do not rely on the origin of the specific spin-dependent tunneling, but do depend on its symmetry.

An immediate consequence of spin-dependence of the tunneling is that the eigenvalue of $\hat{s}_z$ is no longer conserved for tunneling electrons and hence all the projections of the spin fluctuate quantum-mechanically. This opens the question of whether or not these fluctuations result in the appearance of magnetization in the BCS superconductors due to the emergence of spin-polarized Cooper pairs. The answer to this question is the scope of this paper. Although spin-dependent tunneling generates quantum spin fluctuations, we find that for an unbiased system the formation of spin-triplet Cooper pairs is blocked by destructive interference between different quasi-electron and quasi-hole tunneling channels. More precisely, the blockade occurs if the densities of states of quasi-electron and quasi-hole excitations are the same (which is the case if the normal electron density of states is constant across the fermi level). This blockade can, however, be lifted by applying a voltage bias that creates an electrostatically induced imbalance between the tunneling channels.

Without a blockade, two types of spin-dependent tunneling processes for creating spin-triplet Cooper pairs are possible [Fig. 2]. The first type of tunneling involves the **reflection** of one of the electrons of a Cooper pair, which first tunnels from one lead to the other and then back, while flipping its spin. It gives rise to static magnetization in each of the superconducting leads, proportional to the square of the order parameter (energy gap) in that lead. Either a bias voltage or an intrinsic imbalance in the densities of states suffice for creating this static magnetization. The second type of processes is due to **transmission** of a Cooper pair from one superconducting lead to the other by the sequential tunneling of the two paired electrons, with one of them flipping its spin. In the presence of a bias voltage it leads to a time-dependent magnetization and an ensuing oscillatory spin current, both proportional to a product of the order parameters of the two superconductors. Both processes are examples of electrically controlled dissipation-less spintronic phenomena.

*Description of the system —* The Hamiltonian of the junction comprises the terms

$$\mathcal{H} = \mathcal{H}_L + \mathcal{H}_R + \mathcal{H}_{\text{tun}} , \qquad (1)$$

for left and right BCS superconducting leads coupled by spin-dependent tunneling,

$$\mathcal{H}_{\text{tun}} = \sum_{\mathbf{k},\mathbf{p}} \sum_{\sigma,\sigma'} \left( c^\dagger_{\mathbf{k}\sigma} \mathcal{J}_{\mathbf{k}\sigma,\mathbf{p}\sigma'} c_{\mathbf{p}\sigma'} + c^\dagger_{\mathbf{p}\sigma'} \mathcal{J}^*_{\mathbf{k}\sigma,\mathbf{p}\sigma'} c_{\mathbf{k}\sigma} \right)$$
$$\equiv \mathcal{H}_{LR} + \mathcal{H}_{RL} , \qquad (2)$$

where $c_{\mathbf{k}\sigma}$ ($c_{\mathbf{p}\sigma}$) is the annihilation operator of an electron with wave vector $\mathbf{k}$ ($\mathbf{p}$) and spin index $\sigma$. The tunneling amplitudes $\mathcal{J}$ (matrices in spin space) are not specified for the time-being. Different electrochemical potentials, $\mu_L$ and $\mu_R$, are assigned to the quasi-particles in the left and right leads, charging oppositely the superconductors, thus creating charge surface layers at the vicinity of the weak link [1]. These account for a finite bias applied to the junction by adding to the Hamiltonian the terms $-\mu_L N_L - \mu_R N_R$, with $N_{L(R)} = \sum_{\mathbf{k(p)},\sigma} c^\dagger_{\mathbf{k(p)}\sigma} c_{\mathbf{k(p)}\sigma}$ being the number operators. The applied voltage $eV = \mu_L - \mu_R$ is conveniently treated by performing a gauge transformation [2, 3], $U(t) = \exp[-ie\frac{V}{2}(N_L - N_R)t]$, which turns the tunneling Hamiltonian (2) into $\mathcal{H}_{\text{tun}}(t) = \exp[ieVt]\mathcal{H}_{LR} + \exp[-ieVt]\mathcal{H}_{RL}$. At this stage [3] one may replace the tunneling amplitude by

$$\mathcal{J}_{\mathbf{k}\sigma,\mathbf{p}\sigma'}(t) = \exp[i(\phi_0/2 + eVt)]\mathcal{J}_{\mathbf{k}\sigma,\mathbf{p}\sigma'} , \qquad (3)$$

where $\phi_0$ is the phase difference between the order parameters of the two superconducting electrodes.

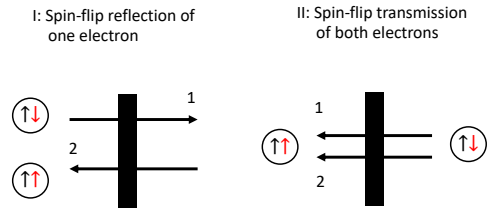

FIG. 2. (color online) Two types of double-tunneling events contributing to the magnetization on the left superconductor. An arrow to the right (left) represents $\mathcal{J}^*$ ($\mathcal{J}$). If, during one of these events, an electron flips its spin, a spin-singlet Cooper pair is converted to a spin-triplet pair, with a finite magnetic moment.

*Magnetization and spin current —* Here we present expressions for the magnetization and the spin current that appear in the left superconducting lead; those pertaining to the right lead are obtained in a similar fashion. Formally, the magnetization $\mathbf{M}_L(t)$ is given by

$$\mathbf{M}_L(t) = (g/2)\mu_B \sum_{\mathbf{k},\sigma,\sigma'} \langle c^\dagger_{\mathbf{k}\sigma}(t) \boldsymbol{\sigma}_{\sigma\sigma'} c_{\mathbf{k}\sigma'}(t) \rangle , \qquad (4)$$

where $\boldsymbol{\sigma}$ is the vector of the Pauli matrices and the angular brackets indicate quantum-mechanical averaging. We use units where $(g/2)\mu_B = 1$, $g$ being the $g$-factor and $\mu_B$ the Bohr magneton, and $\hbar = 1$.

The spin current associated with the left lead is the time derivative of the magnetization,

$$I_L^{\text{spin}}(t) \equiv \dot{\mathbf{M}}_L(t) = \frac{d}{dt}\langle \sum_{\mathbf{k},\sigma,\sigma'} c_{\mathbf{k}\sigma}^\dagger(t) \boldsymbol{\sigma}_{\sigma\sigma'} c_{\mathbf{k}\sigma'}(t) \rangle . \quad (5)$$

The detailed calculation of Eqs. (4) and (5) is relegated to the supplemental material Ref. [15].

The magnetization comprises two distinct contributions, whose physical origins, as discussed above, are quite different,

$$\mathbf{M}_L(t) = \mathbf{M}_{L,\text{dc}} + \mathbf{M}_{L,\text{ac}}(t) . \quad (6)$$

The dc magnetization reads

$$\mathbf{M}_{L,\text{dc}} = \sum_{\mathbf{k},\mathbf{p}} \frac{\Delta_L^2}{4E_k^3} \sum_{\sigma,\sigma'} \boldsymbol{\sigma}_{\sigma'\sigma}[\mathcal{W}_{\sigma\sigma'}^s(\mathbf{k},\mathbf{p}) + \sigma\sigma'\mathcal{W}_{-\sigma'-\sigma}^s(-\mathbf{k},-\mathbf{p})]$$

$$\times \left[ \frac{f_p - f_k}{(E_k - E_p)^2 - (eV)^2}\left(\frac{\xi_p}{E_p}(E_p - E_k) + eV\right) \right.$$
$$\left. + \frac{1 - f_k - f_p}{(E_k + E_p)^2 - (eV)^2}\left(\frac{\xi_p}{E_p}(E_p + E_k) + eV\right) \right], \quad (7)$$

with

$$\mathcal{W}_{\sigma\sigma'}^s(\mathbf{k},\mathbf{p}) = \sum_{\sigma_1} \mathcal{J}_{\mathbf{k}\sigma,\mathbf{p}\sigma_1} \mathcal{J}_{\mathbf{k}\sigma',\mathbf{p}\sigma_1}^* . \quad (8)$$

In Eq. (7) $E_{k(p)} = (\xi_{k(p)}^2 + \Delta_{L(R)}^2)^{1/2}$, and $f_{k(p)} = [\exp(E_{k(p)}/(k_B T)) + 1]^{-1}$ ($\xi_{k(p)}$ is the single-particle energy measured from the common chemical potential of the unbiased device). Note that the static magnetization appears in the superconducting lead even when the other lead is not a superconductor.

Examining the structure of the spin-dependent amplitudes which involves $\mathcal{J}\mathcal{J}^\dagger$ (when the tunneling amplitudes are considered as matrices in spin space), it is seen that the static magnetization indeed results from processes in which one of the electrons of a Cooper pair tunnels from one lead to the other and then back, while flipping its spin. However, though time-reversal symmetry is broken by the bias voltage, this is not enough to ensure a nonzero static magnetization. The most transparent example is that of spin-dependent tunneling due to spin-orbit interaction which by itself conserves time-reversal symmetry. In that case [4]

$$\mathcal{J}_{\mathbf{p}\sigma,\mathbf{k}\sigma'} = \mathcal{J}_{-\mathbf{p}-\sigma,-\mathbf{k}-\sigma'}^* , \quad (9)$$

and consequently the spin-dependent factor is

$$\sum_{\sigma,\sigma'} \boldsymbol{\sigma}_{\sigma\sigma'}\mathcal{W}_{\sigma\sigma'}^s(\mathbf{k},\mathbf{p})(1 + \sigma\sigma')$$
$$= 2\hat{\mathbf{z}}[\mathcal{W}_{\uparrow\uparrow}^s(\mathbf{k},\mathbf{p}) - \mathcal{W}_{\downarrow\downarrow}^s(\mathbf{k},\mathbf{p})] , \quad (10)$$

which vanishes when the spin-orbit coupling is due to the Rashba interaction. Stated differently, the Aharonov-Casher phase [17] accumulated via one tunneling process is cancelled by the accompanying backward time-reversed second tunneling process, so that the spin dependence of the tunneling disappears. Similar to the conductance of two normal metals coupled by spin-orbit active tunneling barrier, that exhibits spin effects only when time-reversal symmetry in the barrier is broken, for instance by a magnetic field [18], a nonzero dc magnetization will be established in the superconducting lead(s) when a Zeeman field (in energy units), $\mathbf{B}$, acts on the electrons passing through the tunneling barrier. In that case $\mathcal{J}_{\mathbf{p}\sigma,\mathbf{k}\sigma'}(\mathbf{B}) = \mathcal{J}_{-\mathbf{p}-\sigma,-\mathbf{k}-\sigma'}^*(-\mathbf{B})$ and the cancellation of the Aharonov-Casher phase disappears.

An explicit example is worked out in the supplemental material [15]. For electrons tunneling through a weak link of length $d$ lying along $\hat{\mathbf{x}}$, in which a Rashba interaction whose pseudo magnetic field is along $\hat{\mathbf{z}}$ and a Zeeman field along $\hat{\mathbf{y}}$ are active, the tunneling amplitude reads

$$\frac{\mathcal{J}(B_y)}{J_0} = \left(-\exp[ik_{\text{so}}d\sigma_z] + \sigma_y \frac{m^* B_y}{k_F^2}[\cos(k_{\text{so}}d) \right.$$
$$\left. - i\frac{\sqrt{k_F^2 + k_{\text{so}}^2}}{k_{\text{so}}}\sin(k_{\text{so}}d)]\right) . \quad (11)$$

Here $J_0$ is the bare tunneling amplitude, $k_{\text{so}}$ is the strength of the Rashba interaction (in momentum units), $k_F$ is the Fermi momentum in the weak link, and a Coulomb blockade of double-electron tunneling was assumed [4, 18]. Exploiting this tunneling amplitude, one finds that the spin-dependent factor in Eq. (7) is $-[8m^* B_y/k_F^2]\cos(k_{\text{so}}d)[\sin(k_{\text{so}}d)\hat{\mathbf{x}} + \cos(k_{\text{so}}d)\hat{\mathbf{y}}]$, proportional to $B_y$.

In contrast, the ac part of the magnetization results from quite different tunneling processes. As shown in Ref. [15], it necessitates that both leads will be superconductors, and has features resembling the Josephson current. It reads

$$\mathbf{M}_{L,\text{ac}}(t) = \sum_{\mathbf{k},\mathbf{p}} \frac{\Delta_L \Delta_R}{2E_k E_p}\text{Re}[\sum_{\sigma,\sigma'} \sigma \boldsymbol{\sigma}_{\sigma\sigma'}\mathcal{W}_{\sigma\sigma'}^a(\mathbf{k},\mathbf{p},t)]$$
$$\times \frac{eV - \xi_k}{E_k^2 - (eV)^2}\left(\frac{(2E_k - E_p)(f_p - f_k)}{(E_k - E_p)^2 - (eV)^2} \right.$$
$$\left. - \frac{(2E_k + E_p)(1 - f_k - f_p)}{(E_k + E_p)^2 - (eV)^2}\right) , \quad (12)$$

where

$$\mathcal{W}_{\sigma\sigma'}^a(\mathbf{k},\mathbf{p},t) = e^{i(\phi_0 + 2eVt)}$$
$$\times \sum_{\sigma''} \sigma'' \mathcal{J}_{\mathbf{k}\sigma',\mathbf{p}\sigma''} \mathcal{J}_{-\mathbf{k}-\sigma,-\mathbf{p}-\sigma''} . \quad (13)$$

$$\mathcal{W}_{\uparrow\downarrow}^a(\mathbf{k},\mathbf{p},t) = e^{i(\phi_0+2eVt)}\left(\mathcal{J}_{\mathbf{k}\downarrow,\mathbf{p}\uparrow}\mathcal{J}_{\mathbf{k}\uparrow,\mathbf{p}\uparrow}^* - \mathcal{J}_{\mathbf{k}\downarrow,\mathbf{p}\downarrow}\mathcal{J}_{\mathbf{k}\uparrow,\mathbf{p}\downarrow}^*\right)$$

The ac oscillations are similar to those of the Josephson current. So is also the appearance of the product

$\mathcal{J}\mathcal{J}$, heralding that the processes leading to the ac magnetization (and the ensuing spin current) result from pair transfer. The spin current through the junction [see Eq. (5)] is obtained from the time derivative of the last term in Eq. (12) [see also Eq. (13)]. For instance, assuming for brevity that the spin factor $\sum_{\sigma,\sigma'} \sigma \boldsymbol{\sigma}_{\sigma\sigma'} \mathcal{W}^a_{\sigma\sigma'}(\mathbf{k}, \mathbf{p}, t)]$ is due solely to a Rashba interaction in the weak link rendering the tunneling amplitudes time-reversal symmetric [Eq. (9)], this factor becomes [15],

$$2\sum_\sigma [\hat{\mathbf{z}} \cos(\phi_0 + 2eVt)\sigma(|\mathcal{J}_{\mathbf{k}\sigma,\mathbf{p}\sigma}|^2 - |\mathcal{J}_{\mathbf{k}\sigma,\mathbf{p}-\sigma}|^2)$$
$$+ 4\sin(\phi_0 + 2eVt)[\hat{\mathbf{y}} \operatorname{Re} \sum_\sigma \sigma \mathcal{J}_{\mathbf{k}-\sigma,\mathbf{p}\sigma} \mathcal{J}^*_{\mathbf{k}\sigma,\mathbf{p}\sigma}$$
$$- \hat{\mathbf{x}} \operatorname{Im} \sum_\sigma \sigma \mathcal{J}_{\mathbf{k}-\sigma,\mathbf{p}\sigma} \mathcal{J}^*_{\mathbf{k}\sigma,\mathbf{p}\sigma}] \ . \tag{14}$$

For a weak link [15] (in the $x-y$ plane) comprising two segments, $\mathbf{r}_L$ and $\mathbf{r}_R$, of equal length, $d/2$, whose angles with the $\mathbf{x}$-direction are $\theta$ and $-\theta$, respectively, one finds that when the electric field creating the Rashba spin-orbit interaction is along $\hat{\mathbf{z}}$, the spin factor becomes

$$-8\sin(\phi_0 + 2eVt)\sin(k_{\mathrm{so}}d)\cos(\theta)\Big(\hat{\mathbf{x}}\sin^2(k_{\mathrm{so}}d/2)\sin(2\theta)$$
$$+ \hat{\mathbf{y}}[\cos^2(k_{\mathrm{so}}d/2) - \sin^2(k_{\mathrm{so}}d/2)\cos(2\theta)]\Big) \ , \tag{15}$$

while when this field is directed along $\hat{\mathbf{y}}$ the spin factor vanishes.

If $k_B T \ll eV \lesssim \Delta_L = \Delta_R = \Delta$ the magnitude of $\mathbf{M}_{L,\mathrm{ac}}(t)$ as given by Eq. (12) can be estimated to be of order $\mu_B(eV/\Delta)(R_Q/R_n)$, where $R_Q = \hbar/e^2 \sim 4\,\mathrm{k\Omega}$ is the resistance quantum and $R_n$ is the resistance of the weak link in the normal state [15]. This corresponds to an induced ac magnetization of order 1-10 mT [15], which could perhaps act as a point source of spin waves that may be detected if a ferromagnetic substrate is used.

By taking the time derivative of Eq. (12), which adds a factor $2eV/\hbar$, we get a Josephson-like spin current carried by Cooper pairs through the weak link. It is interesting to compare the prefactor $I_0^{\mathrm{spin}}$ with the critical current for the ordinary Josephson (charge) current, given as $I_0^{\mathrm{charge}} = (\pi\Delta)/(eR_n)$ by the Ambegaokar-Baratoff formula [19]. One finds that the ratio between these currents, normalized respectively to $\mu_B$ and $e$, is $(I_0^{\mathrm{spin}}/\mu_B)/(I_0^{\mathrm{charge}}/e) \sim (eV/\Delta)^2$ which can be of order 1.

*Discussion* — Non-dissipative electron transport in a superconductor is a property of its single macroscopic quantum ground state (here taken to be the BCS ground state) and is therefore subject to various interference phenomena. A well-known example of such an interference phenomenon is the dc Josephson effect, where a non-dissipative current between two superconductors connected by a weak link is a function of the phase difference between their ground states.

In this paper we have presented a new type of a quantum interference phenomenon, which governs spin-dependent tunneling through a weak link between two bulk superconducting leads. We have shown that a bias on the junction (together with such tunneling) creates spin-triplet Cooper pairs, giving rise to both a Josephson-like ac spin current and a static magnetization in the leads. While static magnetizations can appear in the absence of a bias in superconductors lacking balance between their quasi-electron and quasi-hole states, the spin current necessitates the junction to be biased. As we discuss below, the vanishing of the ac spin current in the absence of a bias voltage can be viewed as the result of destructive interference between different channels of electron tunneling.

Tunneling processes, which make the spins fluctuate and create spin-triplet Cooper pairs, can be viewed as scattering events between different eigenstates of the spin operator $\hat{s}_z$ and can be treated by second-order perturbation in the tunneling Hamiltonian. We have identified two possible processes capable of creating spin-triplet Cooper pairs, one involving the transmission of a Cooper pair from one lead to the other while being converted from a spin-singlet to a spin-triplet Cooper pair. The other represents the reflection of one of the electrons of a Cooper pair, which first tunnels from one lead to the other and then in a second step tunnels back, flipping its spin going either from left to right or from right to left. Here spin-singlet Cooper pairs are converted to being spin-triplet pairs, while staying in the same lead. In the presence of a bias voltage, the transmission-type tunneling gives rise to a time-dependent magnetization and an oscillatory spin current, while the reflection-type leads to a static magnetization. The latter can occur even in an unbiased system, provided that the leads' densities of states are imbalanced, having for instance an excess of quasi-electron states as compared to the quasi-hole ones. We note that if none or both of the paired electrons flip their spin during a transmission event the transferred spin-singlet Cooper pair will contribute to the Josephson charge current, while if during the reflection event the electron does not flip its spin while tunneling back and forth, or flips it twice, no effect will be produced.

In both types of tunneling, transmission and reflection, the system is in a virtual state after the first tunneling event, with one electron in an excited quasi-electron or quasi-hole state in the left lead and the other electron in an excited quasi-electron or quasi-hole state in the right lead. This follows from the Bogoliubov representation of an electron operator in terms of the quasi-electron and quasi-hole ones, $\gamma$, for instance

$$c_{\mathbf{k}\sigma} = u_k \gamma_{\mathbf{k}\sigma} + \sigma v_k \gamma^\dagger_{-\mathbf{k}-\sigma} \ , \tag{16}$$

where $u_k^2 = (1 + \xi_k/E_k)/2$. and $v_k^2 = (1 - \xi_k/E_k)/2$. There are obviously four possibilities (left excitation, right excitation) = (quasi-electron, quasi-electron), (quasi-electron, quasi-hole), (quasi-hole, quasi-hole) and (quasi-hole, quasi-electron) corresponding to four different channels for tunneling. This can be seen by inspecting the perturbation expression [Eq. (14) of Ref. 15] for the left-lead magnetization, $\mathbf{M}_L(t)$,

$$\mathbf{M}_L(t) = -\sum_{\mathbf{k},\sigma,\sigma'} \boldsymbol{\sigma}_{\sigma\sigma'} \int^t dt_1 \int^{t_1} dt_2$$
$$\times \langle [\mathcal{H}_{\text{tun}}(t_2), [\mathcal{H}_{\text{tun}}(t_1), c_{\mathbf{k}\sigma}^\dagger(t) c_{\mathbf{k}\sigma'}(t)]] \rangle , \quad (17)$$

[see Ref. 15 for details].

The total probability amplitude for spin-triplet Cooper pair formation by either transmission or reflection as described above is the sum of the probability amplitudes associated with the four channels. Since these are probability amplitudes rather than probabilities they will, in the language of quantum mechanics, interfere with each other. As seen in the expressions for the magnetization above, it vanishes when the junction is not biased and the quasi-electron and quasi-hole states in the superconductors are balanced, i.e., this interference is then destructive and no spin-triplet Cooper pairs are formed. More precisely, the (quasi-electron, quasi-hole) and (quasi-hole, quasi-electron) channel amplitudes interfere destructively as do the (quasi-electron, quasi-electron) and (quasi-hole, quasi-hole) channel amplitudes. This is the result of the electron-hole symmetry in BCS superconductors and holds for both types of tunneling, transmission and reflection.

The crucial condition for the magnetization to appear is an imbalance of quasi-electron and quasi-hole states. If a voltage bias $V$ is applied to the weak link there will be an electric field between the two superconductors that will break the balance between the effect of these states by shifting the energy of the quasi-hole and quasi-electron states by $eV$. This shift lifts the destructive interference, so that the probability for spin-triplet Cooper-pair formation becomes finite and allows us to view the role of the bias voltage for magnetizing the superconducting leads as a field effect on spin-dependent tunneling. The application of a bias voltage also affects the superconducting phase difference according to the Josephson relation, leading to oscillatory magnetization, which implies the existence of a spin current.

We emphasize that the consequences of Cooper-pair transmission and single-electron reflection of one-half of a Cooper pair for the magnetization of the superconductors are qualitatively different. The transmission of spin-triplet Cooper pairs — when triggered by a bias voltage — resembles the ac Josephson charge-current carried by spin-singlet Cooper pairs. In contrast, the reflection process gives rise to a static (time-independent) magnetization, which can be triggered and tuned electrically by a dc bias voltage. This static magnetization is a ground state property of the superconductors, it exists in the remaining superconductor even if the other is replaced by a normal metal, and it is not related to any charge or spin current in the device. This is in contrast to the nonequilibrium, time-dependent, and nonlinear response of quasi-electrons and quasi-holes in many suggested voltage-biased superconducting hybrid structures containing ferromagnetic elements [10, 20–24]. Both the predicted dc and ac magnetization effects are interesting examples of dissipation-less superconducting spintronic phenomena.

The effect of spin-polarized Cooper-pair tunneling can be enhanced if one of the bulk superconductors is replaced by a small superconducting grain. In this case spin-triplet Cooper pairs injected into the grain can be accumulated over time until a stationary state is reached, where no supercurrent flows and a certain fraction of the Cooper pairs is spin-polarized. The physics would be similar to that of a superconducting magnetic alloy, where paramagnetic impurities give rise to spin-flip scattering [1]. In our case spin-dependent tunneling plays the role of impurity-induced spin scattering, an important difference is that the scattering is not random but well-controlled and electrostatically tunable. As a result, a tunable net magnetization of the grain can be expected.

In the magnetic-alloy case suppression of spin-singlet pairing is controlled by the parameter $(\hbar/\tau_s)/\Delta$, where $\tau_s$ is the spin-flip scattering time. In our case the role of $\tau_s$ is determined by the modulus of the inverse probability amplitude for an electron to tunnel and flip its spin [25], $\hbar J_0^2/[\Gamma|\mathcal{W}_{\uparrow,\downarrow}^a|]$. Here $\Gamma$ is the width of electronic states in the grain due to tunneling when it is in the normal state and $\mathcal{W}_{\uparrow,\downarrow}^a$ is defined in Eq. (13). If the parameter $[\Gamma|\mathcal{W}_{\uparrow,\downarrow}^a|/J_0^2]/\Delta$ can reach a value of order one then a significant fraction of Cooper pairs would be spin-polarized.

Finally, we note that the effect of spin-singlet and spin-triplet Cooper pairs on the superconducting charge- and spin currents are different. When it comes to the charge current only the injection of spin-singlet Cooper pairs into the ground state of a BCS superconductor can be supported. Therefore spin-triplet Cooper pairs are filtered out from contributing to superconducting charge transport. When it comes to the spin current, only spin-triplet Cooper pairs can contribute and spin-singlet Cooper pairs are filtered out. We hence observe an interesting new example of spin-charge separation.

*Acknowledgement* — We acknowledge the hospitality of the PCS at IBS, Daejeon, Korea, where part of this work was supported by IBS funding number (IBS-R024-D1).

**SUPPLEMENTAL MATERIAL FOR**
**"MAGNETIZATION NEAR A CONSTRICTION BETWEEN BCS SUPERCONDUCTORS BY SPIN-DEPENDENT TUNNELING"**

**The Hamiltonian.** The Hamiltonian of the junction comprises the terms

$$\mathcal{H} = \mathcal{H}_L + \mathcal{H}_R + \mathcal{H}_{\text{tun}} - \mu_L \sum_{\mathbf{k},\sigma} c^\dagger_{\mathbf{k}\sigma} c_{\mathbf{k}\sigma} - \mu_R \sum_{\mathbf{p},\sigma} c^\dagger_{\mathbf{p}\sigma} c_{\mathbf{p}\sigma} \ , \tag{S1}$$

where different electrochemical potentials, $\mu_L$ and $\mu_R$, are assigned to the quasi-particles in the left and right leads, charging oppositely the superconductors, thus creating charge surface layers at the vicinity of the weak link. The phenomenon is similar to electron-hole imbalance created by electric current flowing through a superconductor-normal interface, which results in the penetration of an electric field into the superconductors due to the current flow [S1].

In Eq. (S1), $\mathcal{H}_L$ ($\mathcal{H}_R$) is the Hamiltonian of the left (right) superconductor

$$\mathcal{H}_L = \sum_{\mathbf{k},\sigma} \xi_k c^\dagger_{\mathbf{k}\sigma} c_{\mathbf{k}\sigma} - \frac{\Delta_L}{2} \sum_{\mathbf{k},\sigma} \sigma \left( c^\dagger_{\mathbf{k}\sigma} c^\dagger_{-\mathbf{k}-\sigma} + c_{-\mathbf{k}-\sigma} c_{\mathbf{k}\sigma} \right) \ , \tag{S2}$$

with $\Delta_L$ ($\Delta_R$) being the superconducting order parameter in the left (right) lead and $c_{\mathbf{k}\sigma}$ ($c_{\mathbf{p}\sigma}$) denoting the annihilation operator of an electron with wave vector $\mathbf{k}$ ($\mathbf{p}$) and spin index $\sigma$ ($\sigma$ in front of the operators stands for $1, -1$ for $\uparrow, \downarrow$, respectively). The energy $\xi_k$ ($\xi_p$) in the left (right) lead is measured with respect to the common chemical potential of the two superconductors. The (left-lead) Hamiltonian is diagonalized as usual by the Bogolyubov transformation

$$c_{\mathbf{k}\sigma} = u_k \gamma_{\mathbf{k}\sigma} + \sigma v_k \gamma^\dagger_{-\mathbf{k}-\sigma} \ , \quad c^\dagger_{\mathbf{k}\sigma} = u_k \gamma^\dagger_{\mathbf{k}\sigma} + \sigma v_k \gamma_{-\mathbf{k}-\sigma} \ , \tag{S3}$$

with

$$u_k^2 = \frac{1}{2}\left(1 + \frac{\xi_k}{E_k}\right) \ , \quad v_k^2 = \frac{1}{2}\left(1 - \frac{\xi_k}{E_k}\right) \ , \quad u_k v_k = \frac{\Delta_L}{2 E_k} \ , \quad E_k^2 = \xi_k^2 + \Delta_L^2 \ , \tag{S4}$$

to become

$$\mathcal{H}_L = \sum_{\mathbf{k},\sigma} E_k \gamma^\dagger_{\mathbf{k}\sigma} \gamma_{\mathbf{k}\sigma} \ , \tag{S5}$$

with

$$\langle \gamma^\dagger_{\mathbf{k}\sigma}(t) \gamma_{\mathbf{k}\sigma}(t') \rangle = e^{i E_k (t - t')} f(E_k) \ , \quad f(E_k) \equiv f_k = (\exp[\beta E_k] + 1)^{-1} \ . \tag{S6}$$

Analogous expressions pertain for the right-lead Hamiltonian. The tunneling Hamiltonian is

$$\mathcal{H}_{\text{tun}} = \sum_{\mathbf{k},\mathbf{p}} \sum_{\sigma,\sigma'} (c^\dagger_{\mathbf{k}\sigma} \mathcal{J}_{\mathbf{k}\sigma,\mathbf{p}\sigma'} c_{\mathbf{p}\sigma'} + c^\dagger_{\mathbf{p}\sigma'} \mathcal{J}^*_{\mathbf{k}\sigma,\mathbf{p}\sigma'} c_{\mathbf{k}\sigma}) \equiv \mathcal{H}_{LR} + \mathcal{H}_{RL} \ , \tag{S7}$$

with the tunneling amplitudes $\mathcal{J}$ being matrices in spin space.

**Finite bias voltage.** In the presence of an applied voltage $eV = \mu_L - \mu_R$ it is convenient to perform a gauge transformation [S2, S3],

$$U(t) = \exp[-ie\frac{V}{2}(N_L - N_R)t] \ , \quad N_{L(R)} = \sum_{\mathbf{k(p)},\sigma} c^\dagger_{\mathbf{k(p)}\sigma} c_{\mathbf{k(p)}\sigma} \ . \tag{S8}$$

which turns the tunneling Hamiltonian into

$$\mathcal{H}_{\text{tun}}(t) = \exp[ieVt]\mathcal{H}_{LR} + \exp[-ieVt]\mathcal{H}_{RL} \ . \tag{S9}$$

At this stage [S3] one may replace the tunneling amplitude by

$$\mathcal{J}_{\mathbf{k}\sigma,\mathbf{p}\sigma'}(t) = \exp[i(\phi_0 + 2eVt)/2]\mathcal{J}_{\mathbf{k}\sigma,\mathbf{p}\sigma'} \ , \tag{S10}$$

where $\phi_0$ is the phase difference between the order parameters of the two superconducting electrodes.

**The magnetization $\mathbf{M}_L(t)$ (in the left lead).** This quantity is given by

$$\mathbf{M}_L(t) = \sum_{\mathbf{k},\sigma,\sigma'} \langle c^\dagger_{\mathbf{k}\sigma}(t)\boldsymbol{\sigma}_{\sigma\sigma'}c_{\mathbf{k}\sigma'}(t)\rangle \;, \tag{S11}$$

where $\boldsymbol{\sigma}$ is the vector of the Pauli matrices and the angular brackets indicate quantum-mechanical averaging. To second-order perturbation theory (in the tunneling Hamiltonian)

$$\mathbf{M}_L(t) = -\sum_{\mathbf{k},\sigma,\sigma'} \boldsymbol{\sigma}_{\sigma\sigma'} \int^t dt_1 \int^{t_1} dt_2 \langle [\mathcal{H}_{\text{tun}}(t_2), [\mathcal{H}_{\text{tun}}(t_1), c^\dagger_{\mathbf{k}\sigma}(t)c_{\mathbf{k}\sigma'}(t)]]\rangle \;, \tag{S12}$$

where

$$\left[\mathcal{H}_{\text{tun}}(t_1), c^\dagger_{\mathbf{k}\sigma}(t)c_{\mathbf{k}\sigma'}(t)\right] = \sum_{\mathbf{k}_1,\sigma_1} \sum_{\mathbf{p}_1,\sigma'_1} \left(\mathcal{J}_{\mathbf{k}_1\sigma_1,\mathbf{p}_1\sigma'_1}(t_1)\left[c^\dagger_{\mathbf{k}_1\sigma_1}(t_1), c^\dagger_{\mathbf{k}\sigma}(t)c_{\mathbf{k}\sigma'}(t)\right]c_{\mathbf{p}_1\sigma'_1}(t_1)\right.$$
$$\left. + \mathcal{J}^*_{\mathbf{k}_1\sigma_1,\mathbf{p}_1\sigma'_1}(t_1)c^\dagger_{\mathbf{p}_1\sigma'_1}(t_1)\left[c_{\mathbf{k}_1\sigma_1}(t_1), c^\dagger_{\mathbf{k}\sigma}(t)c_{\mathbf{k}\sigma'}(t)\right]\right) \;, \tag{S13}$$

and consequently

$$\mathbf{M}_L(t) = -\sum_{\mathbf{k},\sigma,\sigma'} \boldsymbol{\sigma}_{\sigma\sigma'} \int^t dt_1 \int^{t_1} dt_2 \sum_{\mathbf{p}_1,\sigma'_1} \sum_{\mathbf{p}_2,\sigma'_2} \sum_{\mathbf{k}_1,\sigma_1} \sum_{\mathbf{k}_2,\sigma_2}$$
$$\times \Bigl\{\mathcal{J}_{\mathbf{k}_1\sigma_1,\mathbf{p}_1\sigma'_1}(t_1)\mathcal{J}_{\mathbf{k}_2\sigma_2,\mathbf{p}_2\sigma'_2}(t_2)\Bigl(\langle\left[c^\dagger_{\mathbf{k}_1\sigma_1}(t_1), c^\dagger_{\mathbf{k}\sigma}(t)c_{\mathbf{k}\sigma'}(t)\right]c^\dagger_{\mathbf{k}_2\sigma_2}(t_2)\rangle\langle c_{\mathbf{p}_1\sigma'_1}(t_1)c_{\mathbf{p}_2\sigma'_2}(t_2)\rangle$$
$$- \langle c^\dagger_{\mathbf{k}_2\sigma_2}(t_2)\left[c^\dagger_{\mathbf{k}_1\sigma_1}(t_1), c^\dagger_{\mathbf{k}\sigma}(t)c_{\mathbf{k}\sigma'}(t)\right]\rangle\langle c_{\mathbf{p}_2\sigma'_2}(t_2)c_{\mathbf{p}_1\sigma'_1}(t_1)\rangle\Bigr)$$
$$+ \mathcal{J}^*_{\mathbf{k}_1\sigma_1,\mathbf{p}_1\sigma'_1}(t_1)\mathcal{J}^*_{\mathbf{k}_2\sigma_2,\mathbf{p}_2\sigma'_2}(t_2)\Bigl(\langle\left[c_{\mathbf{k}_1\sigma_1}(t_1), c^\dagger_{\mathbf{k}\sigma}(t)c_{\mathbf{k}\sigma'}(t)\right]c_{\mathbf{k}_2\sigma_2}(t_2)\rangle\langle c^\dagger_{\mathbf{p}_1\sigma'_1}(t_1)c^\dagger_{\mathbf{p}_2\sigma'_2}(t_2)\rangle$$
$$- \langle c_{\mathbf{k}_2\sigma_2}(t_2)\left[c_{\mathbf{k}_1\sigma_1}(t_1), c^\dagger_{\mathbf{k}\sigma}(t)c_{\mathbf{k}\sigma'}(t)\right]\rangle\langle c^\dagger_{\mathbf{p}_2\sigma'_2}(t_2)c^\dagger_{\mathbf{p}_1\sigma'_1}(t_1)\rangle\Bigr)$$
$$+ \mathcal{J}^*_{\mathbf{k}_1\sigma_1,\mathbf{p}_1\sigma'_1}(t_1)\mathcal{J}_{\mathbf{k}_2\sigma_2,\mathbf{p}_2\sigma'_2}(t_2)\Bigl(-\langle\left[c_{\mathbf{k}_1\sigma_1}(t_1), c^\dagger_{\mathbf{k}\sigma}(t)c_{\mathbf{k}\sigma'}(t)\right]c^\dagger_{\mathbf{k}_2\sigma_2}(t_2)\rangle\langle c^\dagger_{\mathbf{p}_1\sigma'_1}(t_1)c_{\mathbf{p}_2\sigma'_2}(t_2)\rangle$$
$$+ \langle c^\dagger_{\mathbf{k}_2\sigma_2}(t_2)\left[c_{\mathbf{k}_1\sigma_1}(t_1), c^\dagger_{\mathbf{k}\sigma}(t)c_{\mathbf{k}\sigma'}(t)\right]\rangle\langle c_{\mathbf{p}_2\sigma'_2}(t_2)c^\dagger_{\mathbf{p}_1\sigma'_1}(t_1)\rangle\Bigr)$$
$$+ \mathcal{J}_{\mathbf{k}_1\sigma_1,\mathbf{p}_1\sigma'_1}(t_1)\mathcal{J}^*_{\mathbf{k}_2\sigma_2,\mathbf{p}_2\sigma'_2}(t_2)\Bigl(-\langle\left[c^\dagger_{\mathbf{k}_1\sigma_1}(t_1), c^\dagger_{\mathbf{k}\sigma}(t)c_{\mathbf{k}\sigma'}(t)\right]c_{\mathbf{k}_2\sigma_2}(t_2)\rangle\langle c_{\mathbf{p}_1\sigma'_1}(t_1)c^\dagger_{\mathbf{p}_2\sigma'_2}(t_2)\rangle$$
$$+ \langle c_{\mathbf{k}_2\sigma_2}(t_2)\left[c^\dagger_{\mathbf{k}_1\sigma_1}(t_1), c^\dagger_{\mathbf{k}\sigma}(t)c_{\mathbf{k}\sigma'}(t)\right]\rangle\langle c^\dagger_{\mathbf{p}_2\sigma'_2}(t_2)c_{\mathbf{p}_1\sigma'_1}(t_1)\rangle\Bigr)\Bigr\} \;. \tag{S14}$$

One notes the crucial difference between the first and second terms on the right hand-side, and the third and fourth ones there. Whereas the first pair of terms are Josephson-like, and oscillate with time [see Eq. (S10)], the second pair whose tunneling amplitudes depend only on the time difference $t_1 - t_2$ will yield a static magnetization.

**The oscillatory magnetization.** We begin with the first term on the right hand-side of Eq. (S14). Using Eqs. (S3) and (S6) one finds that

$$\langle c_{\mathbf{p}_1\sigma'_1}(t_1)c_{\mathbf{p}_2,\sigma'_2}(t_2)\rangle = \delta_{\mathbf{p}_1,-\mathbf{p}_2}\delta_{\sigma'_1,-\sigma'_2}\langle c_{\mathbf{p}_1\sigma'_1}(t_1)c_{-\mathbf{p}_1,-\sigma'_1}(t_2)\rangle \;, \tag{S15}$$

and therefore that first member on the right hand-side of Eq. (S14) becomes

$$\text{first term} = -\sum_{\mathbf{k},\sigma,\sigma'} \boldsymbol{\sigma}_{\sigma\sigma'} \int^t dt_1 \int^{t_1} dt_2 \sum_{\mathbf{p},\sigma''} \sum_{\mathbf{k}_1,\sigma_1} \sum_{\mathbf{k}_2,\sigma_2} \mathcal{J}_{\mathbf{k}_1\sigma_1,\mathbf{p}\sigma''} \mathcal{J}_{\mathbf{k}_2\sigma_2,-\mathbf{p}-\sigma''} e^{i\phi_0+ieV(t_1+t_2)}$$
$$\times \Bigl(\langle\left[c^\dagger_{\mathbf{k}_1\sigma_1}(t_1), c^\dagger_{\mathbf{k}\sigma}(t)c_{\mathbf{k}\sigma'}(t)\right]c^\dagger_{\mathbf{k}_2\sigma_2}(t_2)\rangle\langle c_{\mathbf{p}\sigma''}(t_1)c_{-\mathbf{p}-\sigma''}(t_2)\rangle$$
$$- \langle c^\dagger_{\mathbf{k}_2\sigma_2}(t_2)\left[c^\dagger_{\mathbf{k}_1\sigma_1}(t_1), c^\dagger_{\mathbf{k}\sigma}(t)c_{\mathbf{k}\sigma'}(t)\right]\rangle\langle c_{-\mathbf{p}-\sigma''}(t_2)c_{\mathbf{p}\sigma''}(t_1)\rangle\Bigr) \;. \tag{S16}$$

Here,

$$\langle c_{\mathbf{p}\sigma''}(t_1)c_{-\mathbf{p}-\sigma''}(t_2)\rangle = \sigma''u_p v_p\Bigl(e^{iE_p(t_1-t_2)}f_p - e^{iE_p(t_2-t_1)}(1-f_p)\Bigr) \;,$$
$$\langle c_{-\mathbf{p}-\sigma''}(t_2)c_{\mathbf{p}\sigma''}(t_1)\rangle = u_p v_p\sigma''\Bigl((1-f_p)e^{iE_p(t_1-t_2)} - f_p e^{iE_p(t_2-t_1)}\Bigr) \;, \tag{S17}$$

$$\langle [c^\dagger_{\mathbf{k}_1\sigma_1}(t_1), c^\dagger_{\mathbf{k}\sigma}(t)c_{\mathbf{k}\sigma'}(t)]c^\dagger_{\mathbf{k}_2\sigma_2}(t_2)\rangle = \delta_{\mathbf{k}_1\sigma_1,-\mathbf{k}-\sigma}\delta_{\mathbf{k}_2\sigma_2,\mathbf{k}\sigma'}\langle\{c^\dagger_{\mathbf{k}\sigma}(t), c^\dagger_{-\mathbf{k}-\sigma}(t_1)\}\rangle\langle c_{\mathbf{k}\sigma'}(t)c^\dagger_{\mathbf{k}\sigma'}(t_2)\rangle$$
$$- \delta_{\mathbf{k}_2\sigma_2,-\mathbf{k}-\sigma}\delta_{\mathbf{k}_1\sigma_1,\mathbf{k}\sigma'}\langle\{c^\dagger_{\mathbf{k}\sigma'}(t_1), c_{\mathbf{k}\sigma'}(t)\}\rangle\langle c^\dagger_{\mathbf{k}\sigma}(t)c^\dagger_{-\mathbf{k}-\sigma}(t_2)\rangle$$
$$= \delta_{\mathbf{k}_1\sigma_1,-\mathbf{k}-\sigma}\delta_{\mathbf{k}_2\sigma_2,\mathbf{k}\sigma'}\sigma u_k v_k\Big(e^{iE_k(t_1-t)} - e^{iE_k(t-t_1)}\Big)\Big(u_k^2(1-f_k)e^{iE_k(t_2-t)} + v_k^2 f_k e^{iE_k(t-t_2)}\Big)$$
$$- \delta_{\mathbf{k}_2\sigma_2,-\mathbf{k}-\sigma}\delta_{\mathbf{k}_1\sigma_1,\mathbf{k}\sigma'}\Big(u_k^2 e^{iE_k(t_1-t)} + v_k^2 e^{iE_k(t-t_1)}\Big)\sigma u_k v_k\Big((1-f_k)e^{iE_k(t_2-t)} - f_k e^{iE_k(t-t_2)}\Big), \quad (\text{S18})$$

and

$$\langle c^\dagger_{\mathbf{k}_2\sigma_2}(t_2)[c^\dagger_{\mathbf{k}_1\sigma_1}(t_1), c^\dagger_{\mathbf{k}\sigma}(t)c_{\mathbf{k}\sigma'}(t)]\rangle = \delta_{\mathbf{k}_1\sigma_1,-\mathbf{k}-\sigma}\delta_{\mathbf{k}_2\sigma_2,\mathbf{k}\sigma'}\langle\{c^\dagger_{\mathbf{k}\sigma}(t), c^\dagger_{-\mathbf{k}-\sigma}(t_1)\}\rangle\langle c^\dagger_{\mathbf{k}\sigma'}(t_2)c_{\mathbf{k}\sigma'}(t)\rangle$$
$$- \delta_{\mathbf{k}_2\sigma_2,-\mathbf{k}-\sigma}\delta_{\mathbf{k}_1\sigma_1,\mathbf{k}\sigma'}\langle\{c^\dagger_{\mathbf{k}\sigma'}(t_1), c_{\mathbf{k}\sigma'}(t)\}\rangle\langle c^\dagger_{-\mathbf{k}-\sigma}(t_2)c^\dagger_{\mathbf{k}\sigma}(t)\rangle$$
$$= \delta_{\mathbf{k}_1\sigma_1,-\mathbf{k}-\sigma}\delta_{\mathbf{k}_2\sigma_2,\mathbf{k}\sigma'}\sigma u_k v_k\Big(e^{iE_k(t_1-t)} - e^{iE_k(t-t_1)}\Big)\Big(u_k^2 f_k e^{iE_k(t_2-t)} + v_k^2(1-f_k)e^{iE_k(t-t_2)}\Big)$$
$$- \delta_{\mathbf{k}_2\sigma_2,-\mathbf{k}-\sigma}\delta_{\mathbf{k}_1\sigma_1,\mathbf{k}\sigma'}\Big(u_k^2 e^{iE_k(t_1-t)} + v_k^2 e^{iE_k(t-t_1)}\Big)\sigma u_k v_k\Big(f_k e^{iE_k(t_2-t)} - (1-f_k)e^{iE_k(t-t_2)}\Big), \quad (\text{S19})$$

where all quantum averages are derived using Eqs. (S3) and (S6). As a result, one finds

$$\text{first term} = e^{i\phi_0}\sum_{\mathbf{k},\mathbf{p}} u_k v_k u_p v_p \sum_{\sigma,\sigma'}\sigma\boldsymbol{\sigma}_{\sigma\sigma'}\sum_{\sigma''}\sigma''\mathcal{J}_{\mathbf{k}\sigma',\mathbf{p}\sigma''}\mathcal{J}_{-\mathbf{k}-\sigma,-\mathbf{p}-\sigma''}$$
$$\times\Bigg\{\int^t dt_1\int^{t_1}dt_2 e^{ieV(t_1+t_2)}\Big(e^{iE_k(t_1-t)} - e^{iE_k(t-t_1)}\Big)\Big[(f_p - f_k)\Big(u_k^2 e^{iE_k(t_1-t)}e^{i(E_k-E_p)(t_2-t_1)} + v_k^2 e^{iE_k(t-t_1)}e^{i(E_k-E_p)(t_1-t_2)}\Big)$$
$$- (1 - f_k - f_p)\Big(u_k^2 e^{iE_k(t_1-t)}e^{i(E_k+E_p)(t_2-t_1)} + v_k^2 e^{iE_k(t-t_1)}e^{i(E_k+E_p)(t_1-t_2)}\Big)\Big]$$
$$+ \int^t dt_1\int^{t_1}dt_2 e^{ieV(t_1+t_2)}\Big(u_k^2 e^{iE_k(t_1-t)} + v_k^2 e^{iE_k(t-t_1)}\Big)\Big[(f_p - f_k)\Big(e^{iE_k(t_1-t)}e^{i(E_k-E_p)(t_2-t_1)} - e^{iE_k(t-t_1)}e^{i(E_k-E_p)(t_1-t_2)}\Big)$$
$$- (1 - f_k - f_p)\Big(e^{iE_k(t_1-t)}e^{i(E_k+E_p)(t_2-t_1)} - e^{iE_k(t-t_1)}e^{i(E_k+E_p)(t_1-t_2)}\Big)\Big]\Bigg\}. \quad (\text{S20})$$

Carrying out the $t_2-$integration yields

$$\text{first term} = e^{i\phi_0}\sum_{\mathbf{k},\mathbf{p}} u_k v_k u_p v_p \sum_{\sigma,\sigma'}\sigma\boldsymbol{\sigma}_{\sigma\sigma'}\sum_{\sigma''}\sigma''\mathcal{J}_{\mathbf{k}\sigma',\mathbf{p}\sigma''}\mathcal{J}_{-\mathbf{k}-\sigma,-\mathbf{p}-\sigma''}$$
$$\times\Bigg\{\int^t dt_1 e^{2ieVt_1}\Big(e^{iE_k(t_1-t)} - e^{iE_k(t-t_1)}\Big)\Big[(f_p - f_k)\Big(\frac{u_k^2 e^{iE_k(t_1-t)}}{i(E_k-E_p+eV)} + \frac{v_k^2 e^{iE_k(t-t_1)}}{i(E_p-E_k+eV)}\Big)$$
$$- (1-f_k-f_p)\Big(\frac{u_k^2 e^{iE_k(t_1-t)}}{i(E_k+E_p+eV)} + \frac{v_k^2 e^{iE_k(t-t_1)}}{i(-E_p-E_k+eV)}\Big)\Big]$$
$$+ \int^t dt_1 e^{2ieVt_1}\Big(u_k^2 e^{iE_k(t_1-t)} + v_k^2 e^{iE_k(t-t_1)}\Big)\Big[(f_p-f_k)\Big(\frac{e^{iE_k(t_1-t)}}{i(E_k-E_p+eV)} - \frac{e^{iE_k(t-t_1)}}{i(E_p-E_k+eV)}\Big)$$
$$- (1-f_k-f_p)\Big(\frac{e^{iE_k(t_1-t)}}{i(E_k+E_p+eV)} - \frac{e^{iE_k(t-t_1)}}{i(-E_p-E_k+eV)}\Big)\Big]\Bigg\}. \quad (\text{S21})$$

The integration over $t_1$ results in

$$\text{first term} = \sum_{\mathbf{k},\mathbf{p}}\frac{\Delta_L\Delta_R}{4E_k E_p}e^{i\phi_0+2ieVt}\sum_{\sigma,\sigma'}\sigma\boldsymbol{\sigma}_{\sigma\sigma'}\sum_{\sigma''}\sigma''\mathcal{J}_{\mathbf{k}\sigma',\mathbf{p}\sigma''}\mathcal{J}_{-\mathbf{k}-\sigma,-\mathbf{p}-\sigma''}$$
$$\times\Bigg((f_p-f_k)\frac{(2E_k-E_p)(eV-\xi_k)}{[E_k^2-(eV)^2][(E_k-E_p)^2-(eV)^2]} - (1-f_k-f_p)\frac{(2E_k+E_p)(eV-\xi_k)}{[E_k^2-(eV)^2][(E_k+E_p)^2-(eV)^2]}\Bigg). \quad (\text{S22})$$

Te second member on the right hand-side of Eq. (S14) is treated similarly. The sum of the two,

first term + second term, is the ac magnetization created in the left superconductor, $\mathbf{M}_{L,\mathrm{ac}}(t)$,

$$\mathbf{M}_{L,\mathrm{ac}}(t) = \sum_{\mathbf{k},\mathbf{p}} \frac{\Delta_L \Delta_R}{4E_k E_p} \left( e^{i\phi_0 + 2ieVt} \sum_{\sigma,\sigma'} \sigma \boldsymbol{\sigma}_{\sigma\sigma'} \sum_{\sigma''} \sigma'' \mathcal{J}_{\mathbf{k}\sigma',\mathbf{p}\sigma''} \mathcal{J}_{-\mathbf{k}-\sigma,-\mathbf{p}-\sigma''} + \mathrm{c.c.} \right)$$
$$\times \left( (f_p - f_k) \frac{(2E_k - E_p)(eV - \xi_k)}{[E_k^2 - (eV)^2][(E_k - E_p)^2 - (eV)^2]} - (1 - f_k - f_p) \frac{(2E_k + E_p)(eV - \xi_k)}{[E_k^2 - (eV)^2][(E_k + E_p)^2 - (eV)^2]} \right) . \quad (S23)$$

The ac magnetization comprises two factors: the terms in the first circular brackets specify the spin dependence of the ac magnetization while the terms in the second circular brackets reflect the criteria required for its appearance: the ac magnetization appears either when the junction is biased by a voltage $V$, which leads to an electrostatically-induced imbalance in the electron-hole contributions to the spin-dependent tunneling, or when the density of states (in this case, of the left lead) reflects an imbalance between the quasi-electron and quasi-hole states. This latter is manifested by the appearance of the $\xi_k$ factors in the numerators. When the integration over the momenta is turned into an integration over $\xi_k$, i.e., $\sum_{\mathbf{k}} \to \int d\xi_k \mathcal{N}(\xi_k)$ ($\mathcal{N}$ being the density of states) these terms will disappear unless the density of states has also an odd dependence on $\xi_k$ due to an imbalance between the quasi-particle and quasi-hole states [S1].

The spin-dependence of the tunneling amplitudes may be due to spin-orbit interactions in the weak link. Since this interaction is time-reversal symmetric, it follows that [S4]

$$\mathcal{J}_{-\mathbf{k}-\sigma,-\mathbf{p}-\sigma''} = \mathcal{J}^*_{\mathbf{k}\sigma,\mathbf{p}\sigma''} . \quad (S24)$$

In that case,

$$e^{i\phi_0 + 2ieVt} \sum_{\sigma,\sigma'} \sigma \boldsymbol{\sigma}_{\sigma\sigma'} \sum_{\sigma''} \sigma'' \mathcal{J}_{\mathbf{k}\sigma',\mathbf{p}\sigma''} \mathcal{J}_{-\mathbf{k}-\sigma,-\mathbf{p}-\sigma''} + \mathrm{c.c.}$$
$$= 2\hat{\mathbf{z}} \cos(\phi_0 + 2eVt)(|\mathcal{J}_{\mathbf{k}\uparrow,\mathbf{p}\uparrow}|^2 - |\mathcal{J}_{\mathbf{k}\downarrow,\mathbf{p}\downarrow}|^2 - |\mathcal{J}_{\mathbf{k}\uparrow,\mathbf{p}\downarrow}|^2 + |\mathcal{J}_{\mathbf{k}\downarrow,\mathbf{p}\uparrow}|^2)$$
$$+ 2i\hat{\mathbf{x}} \sin(\phi_0 + 2eVt)(\mathcal{J}_{\mathbf{k}\downarrow,\mathbf{p}\uparrow}\mathcal{J}^*_{\mathbf{k}\uparrow,\mathbf{p}\uparrow} + \mathcal{J}_{\mathbf{k}\uparrow,\mathbf{p}\downarrow}\mathcal{J}^*_{\mathbf{k}\downarrow,\mathbf{p}\downarrow} - \mathcal{J}_{\mathbf{k}\uparrow,\mathbf{p}\uparrow}\mathcal{J}^*_{\mathbf{k}\downarrow,\mathbf{p}\uparrow} - \mathcal{J}_{\mathbf{k}\downarrow,\mathbf{p}\downarrow}\mathcal{J}^*_{\mathbf{k}\uparrow,\mathbf{p}\downarrow})$$
$$+ 2\hat{\mathbf{y}} \sin(\phi_0 + 2eVt)(\mathcal{J}_{\mathbf{k}\downarrow,\mathbf{p}\uparrow}\mathcal{J}^*_{\mathbf{k}\uparrow,\mathbf{p}\uparrow} - \mathcal{J}_{\mathbf{k}\downarrow,\mathbf{p}\downarrow}\mathcal{J}^*_{\mathbf{k}\uparrow,\mathbf{p}\downarrow} + \mathcal{J}_{\mathbf{k}\uparrow,\mathbf{p}\uparrow}\mathcal{J}^*_{\mathbf{k}\downarrow,\mathbf{p}\uparrow} - \mathcal{J}^*_{\mathbf{k}\downarrow,\mathbf{p}\downarrow}\mathcal{J}_{\mathbf{k}\uparrow,\mathbf{p}\downarrow}) . \quad (S25)$$

As an example we consider the two scenarios described in Ref. S4 (see the supplemental material of that reference) where a weak link lying in the $X-Y$ plane, is made of two segments, $\mathbf{r}_L$ and $\mathbf{r}_R$, of equal length, $d/2$, whose angles with the $\mathbf{x}$-direction are $\theta$ and $-\theta$, respectively. Denoting the direction of the electric field creating the Rashba spin-orbit interaction (of strength $k_{\mathrm{so}}$ in momentum units) by $\hat{\mathbf{n}}$ one finds that for $\hat{\mathbf{n}} \| \hat{\mathbf{z}}$

$$\mathcal{J}/J_0 = \cos^2(k_{\mathrm{so}}d/2) - \sin^2(k_{\mathrm{so}}d/2)\cos(2\theta) + i\boldsymbol{\sigma} \cdot [\hat{\mathbf{y}}\sin(k_{\mathrm{so}}d)\cos(\theta) + \hat{\mathbf{z}}\sin^2(k_{\mathrm{so}}d/2)\sin(2\theta)] , \quad (S26)$$

while for $\hat{\mathbf{n}}$ along $\hat{\mathbf{y}}$

$$\mathcal{J}/J_0 = \cos[k_{\mathrm{so}}d\cos(\theta)] - i\boldsymbol{\sigma} \cdot \hat{\mathbf{z}}\sin[k_{\mathrm{so}}d\cos(\theta)] . \quad (S27)$$

($J_0$ is the bare tunneling amplitude.) It is easily seen that in the second case all components in Eq. (S25) vanish. In the first scenario Eq. (S25) yields

$$-8\sin(\phi_0 + 2eVt)\sin(k_{\mathrm{so}}d)\cos(\theta)\left(\hat{\mathbf{x}}\sin^2(k_{\mathrm{so}}d/2)\sin(2\theta) + \hat{\mathbf{y}}[\cos^2(k_{\mathrm{so}}d/2) - \sin^2(k_{\mathrm{so}}d/2)\cos(2\theta)]\right) . \quad (S28)$$

**Estimate of the magnetization magnitude.** Converting the sum over momenta $\mathbf{k}$ to an integral we get

$$\sum_{\mathbf{k}} = \frac{V_L}{(2\pi)^3} \frac{4\pi}{3} \int_0^\infty k^2 dk = \frac{V_L}{6\pi^2} \int_{-\mu}^\infty \frac{mk}{\hbar^2} d\left(\frac{\hbar^2 k^2}{2m} - \mu\right) \approx \frac{V_L k_F^3}{12\pi^2 E_F} \int_{-\infty}^\infty d\xi_k \equiv \rho_L \int_{-\infty}^\infty d\xi_{\mathbf{k}} .$$

Note that $\rho_L$ is the density of states per unit energy (so its dimension is inverse energy) and is proportional to the volume $V_L$ of the left lead. Similarly,

$$\sum_{\mathbf{p}} \approx \frac{V_R k_F^3}{12\pi^2 E_F} \int_{-\infty}^\infty d\xi_p \equiv \rho_R \int_{-\infty}^\infty d\xi_{\mathbf{p}} .$$

The non-spin-dependent part of Eq. (S23) for $\mathbf{M}_{L,\mathrm{ac}}(t)$, i.e. excluding the factors within in the first large round parentheses divided by $\mathcal{J}_0^2$ (the magnitude of the tunneling amplitude), is (multiplied by the magnetic moment of an electron spin, $-(g/2)\mu_B \approx -\mu_B$ and assuming constant densities of states, $\Delta_L = \Delta_R = \Delta$, and $eV, k_B T \ll \Delta$)

$$\mu_B(eV) \sum_{\mathbf{k},\mathbf{p}} \frac{\Delta^2}{4E_k^3 E_p} J_0^2 \frac{(2E_k + E_p)}{(E_k + E_p)^2} = \mu_B \frac{eV}{4} \rho_L \rho_R J_0^2 \int_0^\infty d\xi_k \int_0^\infty d\xi_p \frac{\Delta^2}{E_k^3 E_p} \frac{(2E_k + E_p)}{(E_k + E_p)^2} \ . \tag{S29}$$

The integration over $\xi_{\mathbf{k}(\mathbf{p})}$ will only contribute over a range $\Delta$ so the result is approximately (keeping the numerical coefficient for now)

$$\mu_B \frac{3}{16} \frac{eV}{\Delta} \left(\rho_L \rho_R \mathcal{J}_0^2\right) \approx \mu_B \frac{eV}{\Delta} \frac{R_Q}{R_n} \ , \tag{S30}$$

where $R_Q = \hbar/e^2$ (about 4 k$\Omega$) is the resistance quantum and $R_n = \hbar/(e^2 \rho_L \rho_R \mathcal{J}_0^2)$ is the normal-state resistance of the weak link.

In SI units the magnetic **B**-field (the magnetic flux density) generated by a magnetization field **M** is $\mathbf{B} = \mu_0 \mathbf{M}$. Here $\mu_0 \approx 1.3 \times 10^{-6}$ N/A$^2$ is the magnetic permeability of vacuum and **M** is the magnetization per unit volume. Let us therefore calculate the **B**-field generated by a magnetization $\mu_B$ contained in a (small) volume $V_0$. The value of the Bohr magneton is $\mu_B \approx 9.3 \times 10^{-24}$ J/T so therefore

$$\mathbf{B} = \left(\frac{\mathrm{m}^3}{V_0}\right) \mu_0 \mu_B \, \mathrm{m}^{-3} \approx \left(\frac{\mathrm{m}^3}{V_0}\right) 1.2 \times 10^{-29} \frac{\mathrm{NJ}}{\mathrm{m}^3 \mathrm{A}^2 \mathrm{T}} = \left(\frac{\mathrm{m}^3}{V_0}\right) 1.2 \times 10^{-29} \, \mathrm{T}$$

where we have used that T=N/(Am) and J=Nm, For a volume of $1\,\mathrm{\AA}^3 = 10^{-30}\,\mathrm{m}^3$ we get 1 T (as expected for a ferromagnet) while for a half-sphere of radius 1 nm we get about 10 mT. Taking into account the prefactors above it is reasonable to estimate the field as 1-10 mT.

Considering the spin current we need to add a factor $(2eV/\hbar)$ to get a prefactor

$$I_0^{\mathrm{spin}} = \mu_B \frac{2(eV)^2}{\hbar \Delta} \frac{R_Q}{R_n}; \quad I_0^{\mathrm{spin}}/\mu_B = \frac{2(eV)^2}{e^2 \Delta R_n} \ ,$$

which can be compared to the critical current, $I_0^{\mathrm{charge}}$, of the ordinary Josephson effect, where

$$I_0^{\mathrm{charge}}/e = \frac{\pi \Delta}{e^2 R_n} \ ,$$

so that

$$\frac{I_0^{\mathrm{spin}}/\mu_B}{I_0^{\mathrm{charge}}/e} = \frac{2}{\pi} \left(\frac{eV}{\Delta}\right)^2 \ .$$

It seems reasonable to assume $R_Q/R_n \sim 0.1$ and $eV \sim \Delta$. In our model the induced magnetization appears over the entire volume of the left lead. In reality it is concentrated close to the weak link orifice.

**The static magnetization.** The third term on the right hand-side of Eq. (S14) is treated as follows. Using [see Eqs. (S3) and (S6)]

$$\langle c^\dagger_{\mathbf{p}_1 \sigma_1'}(t_1) c_{\mathbf{p}_2 \sigma_2'}(t_2) \rangle = \delta_{\mathbf{p}_1 \sigma_1', \mathbf{p}\sigma''} \delta_{\mathbf{p}_2 \sigma_2', \mathbf{p}\sigma''} \left(u_p^2 f_p e^{iE_p(t_1-t_2)} + v_p^2 (1-f_p) e^{iE_p(t_2-t_1)}\right) \ ,$$
$$\langle c_{\mathbf{p}_2 \sigma_2'}(t_2) c^\dagger_{\mathbf{p}_1 \sigma_1'}(t_1) \rangle = \delta_{\mathbf{p}_1 \sigma_1', \mathbf{p}\sigma''} \delta_{\mathbf{p}_2 \sigma_2', \mathbf{p}\sigma''} \left(u_p^2 (1-f_p) e^{iE_p(t_1-t_2)} + v_p^2 f_p e^{iE_p(t_2-t_1)}\right) \ , \tag{S31}$$

$$\langle \left[c_{\mathbf{k}_1 \sigma_1}(t_1), c^\dagger_{\mathbf{k}\sigma}(t) c_{\mathbf{k}\sigma'}(t)\right] c^\dagger_{\mathbf{k}_2 \sigma_2}(t_2) \rangle = \delta_{\mathbf{k}\sigma, \mathbf{k}_1 \sigma_1} \delta_{\mathbf{k}\sigma', \mathbf{k}_2 \sigma_2} \langle \{c_{\mathbf{k}\sigma}(t_1), c^\dagger_{\mathbf{k}\sigma}(t)\} \rangle \langle c_{\mathbf{k}\sigma'}(t) c^\dagger_{\mathbf{k}\sigma'}(t_2) \rangle$$
$$- \delta_{-\mathbf{k}-\sigma', \mathbf{k}_1 \sigma_1} \delta_{-\mathbf{k}-\sigma, \mathbf{k}_2 \sigma_2} \langle \{c_{\mathbf{k}\sigma'}(t), c_{-\mathbf{k}-\sigma'}(t_1)\} \rangle \langle c^\dagger_{\mathbf{k}\sigma}(t) c^\dagger_{-\mathbf{k}-\sigma}(t_2) \rangle$$
$$= \delta_{\mathbf{k}\sigma, \mathbf{k}_1 \sigma_1} \delta_{\mathbf{k}\sigma', \mathbf{k}_2 \sigma_2} \left(u_k^2 e^{iE_k(t-t_1)} + v_k^2 e^{iE_k(t_1-t)}\right) \left(u_k^2 (1-f_k) e^{iE_k(t_2-t)} + v_k^2 f_k e^{iE_k(t-t_2)}\right)$$
$$- \delta_{-\mathbf{k}-\sigma', \mathbf{k}_1 \sigma_1} \delta_{-\mathbf{k}-\sigma, \mathbf{k}_2 \sigma_2} u_k^2 v_k^2 \sigma \sigma' \left(e^{iE_k(t-t_1)} - e^{iE_k(t_1-t)}\right) \left((1-f_k) e^{iE_k(t_2-t)} - f_k e^{iE_k(t-t_2)}\right) \ , \tag{S32}$$

and

$$\begin{aligned}
\langle c^\dagger_{\mathbf{k}_2\sigma_2}(t_2)[c_{\mathbf{k}_1\sigma_1}(t_1), c^\dagger_{\mathbf{k}\sigma}(t)c_{\mathbf{k}\sigma'}(t)]\rangle &= \delta_{\mathbf{k}\sigma,\mathbf{k}_1\sigma_1}\delta_{\mathbf{k}\sigma',\mathbf{k}_2\sigma_2}\langle\{c_{\mathbf{k}\sigma}(t_1), c^\dagger_{\mathbf{k}\sigma}(t)\}\rangle\langle c^\dagger_{\mathbf{k}\sigma'}(t_2)c_{\mathbf{k}\sigma'}(t)\rangle \\
&\quad - \delta_{-\mathbf{k}-\sigma',\mathbf{k}_1\sigma_1}\delta_{-\mathbf{k}-\sigma,\mathbf{k}_2\sigma_2}\langle\{c_{\mathbf{k}\sigma'}(t), c_{-\mathbf{k}-\sigma'}(t_1)\}\rangle\langle c^\dagger_{-\mathbf{k}-\sigma}(t_2)c^\dagger_{\mathbf{k}\sigma}(t)\rangle \\
&= \delta_{\mathbf{k}\sigma,\mathbf{k}_1\sigma_1}\delta_{\mathbf{k}\sigma',\mathbf{k}_2\sigma_2}\left(u_k^2 e^{iE_k(t-t_1)} + v_k^2 e^{iE_k(t_1-t)}\right)\left(u_k^2 f_k e^{iE_k(t_2-t)} + v_k^2(1-f_k)e^{iE_k(t-t_2)}\right) \\
&\quad - \delta_{-\mathbf{k}-\sigma',\mathbf{k}_1\sigma_1}\delta_{-\mathbf{k}-\sigma,\mathbf{k}_2\sigma_2} u_k^2 v_k^2 \sigma\sigma'\left(e^{iE_k(t-t_1)} - e^{iE_k(t_1-t)}\right)\left(f_k e^{iE_k(t_2-t)} - (1-f_k)e^{iE_k(t-t_2)}\right),
\end{aligned} \quad \text{(S33)}$$

this term becomes

$$\begin{aligned}
\text{third term} &= -\sum_{\mathbf{k},\mathbf{p}}\sum_{\sigma,\sigma'}\boldsymbol{\sigma}_{\sigma\sigma'}\sum_{\sigma''}\mathcal{J}^*_{\mathbf{k}\sigma,\mathbf{p}\sigma''}\mathcal{J}_{\mathbf{k}\sigma',\mathbf{p}\sigma''}\int^t dt_1\int^{t_1} dt_2 e^{ieV(t_2-t_1)}\left(u_k^2 e^{iE_k(t-t_1)} + v_k^2 e^{iE_k(t_1-t)}\right) \\
&\quad \times\left[\left(u_k^2 u_p^2 e^{iE_k(t_1-t)}e^{i(E_k-E_p)(t_2-t_1)} - v_k^2 v_p^2 e^{iE_k(t-t_1)}e^{i(E_k-E_p)(t_1-t_2)}\right)(f_k-f_p)\right. \\
&\quad \left. + \left(v_k^2 u_p^2 e^{iE_k(t-t_1)}e^{i(E_k+E_p)(t_1-t_2)} - u_k^2 v_p^2 e^{iE_k(t_1-t)}e^{i(E_k+E_p)(t_2-t_1)}\right)(1-f_k-f_p)\right] \\
&\quad + \sum_{\mathbf{k},\mathbf{p}}\sum_{\sigma,\sigma'}\boldsymbol{\sigma}_{\sigma\sigma'}\sigma\sigma'\sum_{\sigma''}\mathcal{J}^*_{-\mathbf{k}-\sigma',\mathbf{p}\sigma''}\mathcal{J}_{-\mathbf{k}-\sigma,\mathbf{p}\sigma''}\int^t dt_1\int^{t_1} dt_2 e^{ieV(t_2-t_1)}u_k^2 v_k^2\left(e^{iE_k(t-t_1)}-e^{iE_k(t_1-t)}\right) \\
&\quad \times\left[\left(u_p^2 e^{iE_k(t_1-t)}e^{i(E_k-E_p)(t_2-t_1)} + v_p^2 e^{iE_k(t-t_1)}e^{i(E_k-E_p)(t_1-t_2)}\right)(f_k-f_p)\right. \\
&\quad \left. - \left(u_p^2 e^{iE_k(t-t_1)}e^{i(E_k+E_p)(t_1-t_2)} + v_p^2 e^{iE_k(t_1-t)}e^{i(E_k+E_p)(t_2-t_1)}\right)(1-f_k-f_p)\right].
\end{aligned} \quad \text{(S34)}$$

Treating the fourth term on the right hand-side of Eq. (S14) in a similar fashion yields

$$\begin{aligned}
\text{fourth term} &= \sum_{\mathbf{k},\mathbf{p}}\sum_{\sigma,\sigma'}\boldsymbol{\sigma}_{\sigma\sigma'}\sigma\sigma'\sum_{\sigma''}\mathcal{J}^*_{-\mathbf{k}-\sigma',\mathbf{p}\sigma''}\mathcal{J}_{-\mathbf{k}-\sigma,\mathbf{p}\sigma''}\int^t dt_1\int^{t_1} dt_2 e^{ieV(t_1-t_2)}u_k^2 v_k^2\left(e^{iE_k(t_1-t)}-e^{iE_k(t-t_1)}\right) \\
&\quad \times\left[\left(u_p^2 e^{iE_k(t-t_1)}e^{i(E_k-E_p)(t_1-t_2)} + v_p^2 e^{iE_k(t_1-t)}e^{i(E_k-E_p)(t_2-t_1)}\right)(f_k-f_p)\right. \\
&\quad \left. - \left(u_p^2 e^{iE_k(t_1-t)}e^{i(E_k+E_p)(t_2-t_1)} + v_p^2 e^{iE_k(t-t_1)}e^{i(E_k+E_p)(t_1-t_2)}\right)(1-f_k-f_p)\right] \\
&\quad - \sum_{\mathbf{k},\mathbf{p}}\sum_{\sigma,\sigma'}\boldsymbol{\sigma}_{\sigma\sigma'}\sum_{\sigma''}\mathcal{J}_{\mathbf{k}\sigma',\mathbf{p}\sigma''}\mathcal{J}^*_{\mathbf{k}\sigma,\mathbf{p}\sigma''}\int^t dt_1\int^{t_1} dt_2 e^{ieV(t_1-t_2)}\left(u_k^2 e^{iE_k(t_1-t)} + v_k^2 e^{iE_k(t-t_1)}\right) \\
&\quad \times\left[\left(u_k^2 u_p^2 e^{iE_k(t-t_1)}e^{i(E_k-E_p)(t_1-t_2)} - v_k^2 v_p^2 e^{iE_k(t_1-t)}e^{i(E_k-E_p)(t_2-t_1)}\right)(f_k-f_p)\right. \\
&\quad \left. + \left(v_k^2 u_p^2 e^{iE_k(t_1-t)}e^{i(E_k+E_p)(t_2-t_1)} - u_k^2 v_p^2 e^{iE_k(t-t_1)}e^{i(E_k+E_p)(t_1-t_2)}\right)(1-f_k-f_p)\right].
\end{aligned} \quad \text{(S35)}$$

Consequently, combining Eqs. (S34) and (S35), one finds

$$\begin{aligned}
\text{third term} + \text{fourth term} &= -2\sum_{\mathbf{k},\mathbf{p}}\sum_{\sigma,\sigma'}\boldsymbol{\sigma}_{\sigma\sigma'}\sigma\sigma'\sum_{\sigma''}\mathcal{J}^*_{-\mathbf{k}-\sigma',\mathbf{p}\sigma''}\mathcal{J}_{-\mathbf{k}-\sigma,\mathbf{p}\sigma''} \\
&\quad \times\text{Re}\Big\{\int^t dt_1\int^{t_1} dt_2 e^{ieV(t_1-t_2)}u_k^2 v_k^2\left(e^{iE_k(t_1-t)}-e^{iE_k(t-t_1)}\right) \\
&\quad \times\left[\left(u_p^2 e^{iE_k(t-t_1)}e^{i(E_k-E_p)(t_1-t_2)} + v_p^2 e^{iE_k(t_1-t)}e^{i(E_k-E_p)(t_2-t_1)}\right)(f_p-f_k)\right. \\
&\quad \left. + \left(v_p^2 e^{iE_k(t-t_1)}e^{i(E_k+E_p)(t_1-t_2)} + u_p^2 e^{iE_k(t_1-t)}e^{i(E_k+E_p)(t_2-t_1)}\right)(1-f_k-f_p)\right]\Big\} \\
&\quad + 2\sum_{\mathbf{k},\mathbf{p}}\sum_{\sigma,\sigma'}\boldsymbol{\sigma}_{\sigma\sigma'}\sum_{\sigma''}\mathcal{J}_{\mathbf{k}\sigma',\mathbf{p}\sigma''}\mathcal{J}^*_{\mathbf{k}\sigma,\mathbf{p}\sigma''}\text{Re}\Big\{\int^t dt_1\int^{t_1} dt_2 e^{ieV(t_1-t_2)}\left(u_k^2 e^{iE_k(t_1-t)} + v_k^2 e^{iE_k(t-t_1)}\right) \\
&\quad \times\left[\left(u_k^2 u_p^2 e^{iE_k(t-t_1)}e^{i(E_k-E_p)(t_1-t_2)} - v_k^2 v_p^2 e^{iE_k(t_1-t)}e^{i(E_k-E_p)(t_2-t_1)}\right)(f_p-f_k)\right. \\
&\quad \left. + \left(u_k^2 v_p^2 e^{iE_k(t-t_1)}e^{i(E_k+E_p)(t_1-t_2)} - v_k^2 u_p^2 e^{iE_k(t_1-t)}e^{i(E_k+E_p)(t_2-t_1)}\right)(1-f_k-f_p)\right]\Big\}.
\end{aligned} \quad \text{(S36)}$$

Integrating over $t_2$ gives

$$\text{third term + fourth term} = -2\sum_{\mathbf{k},\mathbf{p}}\sum_{\sigma,\sigma'}\boldsymbol{\sigma}_{\sigma\sigma'}\sigma\sigma'\sum_{\sigma''}\mathcal{J}^*_{-\mathbf{k}-\sigma',\mathbf{p}\sigma''}\mathcal{J}_{-\mathbf{k}-\sigma,\mathbf{p}\sigma''}\text{Re}\Big\{\int^t dt_1 u_k^2 v_k^2 \Big(e^{iE_k(t_1-t)} - e^{iE_k(t-t_1)}\Big)$$

$$\times\Big[\Big(\frac{u_p^2 e^{iE_k(t-t_1)}}{i(E_p - E_k - eV)} + \frac{v_p^2 e^{iE_k(t_1-t)}}{i(E_k - E_p - eV)}\Big)(f_p - f_k) + \Big(\frac{-v_p^2 e^{iE_k(t-t_1)}}{i(E_k + E_p + eV)} + \frac{u_p^2 e^{iE_k(t_1-t)}}{i(E_k + E_p - eV)}\Big)(1 - f_k - f_p)\Big]\Big\}$$

$$+2\sum_{\mathbf{k},\mathbf{p}}\sum_{\sigma,\sigma'}\boldsymbol{\sigma}_{\sigma\sigma'}\sum_{\sigma''}\mathcal{J}_{\mathbf{k}\sigma',\mathbf{p}\sigma''}\mathcal{J}^*_{\mathbf{k}\sigma,\mathbf{p}\sigma''}\text{Re}\Big\{\int^t dt_1 \Big(u_k^2 e^{iE_k(t_1-t)} + v_k^2 e^{iE_k(t-t_1)}\Big)$$

$$\times\Big[\Big(\frac{u_k^2 u_p^2 e^{iE_k(t-t_1)}}{i(E_p - E_k - eV)} - \frac{v_k^2 v_p^2 e^{iE_k(t_1-t)}}{i(E_k - E_p - eV)}\Big)(f_p - f_k) + \Big(\frac{-u_k^2 v_p^2 e^{iE_k(t-t_1)}}{i(E_k + E_p + eV)} - \frac{v_k^2 u_p^2 e^{iE_k(t_1-t)}}{i(E_k + E_p - eV)}\Big)(1 - f_k - f_p)\Big]\Big\}\ . \tag{S37}$$

Note that half of the terms disappear (since they are purely imaginary) and the rest give

$$\text{third + fourth terms} = -2\sum_{\mathbf{k},\mathbf{p}} u_k^2 v_k^2 \sum_{\sigma,\sigma'}\boldsymbol{\sigma}_{\sigma\sigma'}\sigma\sigma'\sum_{\sigma''}\mathcal{J}^*_{-\mathbf{k}-\sigma',\mathbf{p}\sigma''}\mathcal{J}_{-\mathbf{k}-\sigma,\mathbf{p}\sigma''}$$

$$\times\text{Re}\Big\{\int^t dt_1\Big[\Big(\frac{-u_p^2 e^{2iE_k(t-t_1)}}{i(E_p - E_k - eV)} + \frac{v_p^2 e^{2iE_k(t_1-t)}}{i(E_k - E_p - eV)}\Big)(f_p - f_k) + \Big(\frac{v_p^2 e^{2iE_k(t-t_1)}}{i(E_k + E_p + eV)} + \frac{u_p^2 e^{2iE_k(t_1-t)}}{i(E_k + E_p - eV)}\Big)(1 - f_k - f_p)\Big]\Big\}$$

$$+2\sum_{\mathbf{k},\mathbf{p}} u_k^2 v_k^2 \sum_{\sigma,\sigma'}\boldsymbol{\sigma}_{\sigma\sigma'}\sum_{\sigma''}\mathcal{J}_{\mathbf{k}\sigma',\mathbf{p}\sigma''}\mathcal{J}^*_{\mathbf{k}\sigma,\mathbf{p}\sigma''}\text{Re}\Big\{\int^t dt_1\Big[\Big(\frac{u_p^2 e^{2iE_k(t-t_1)}}{i(E_p - E_k - eV)} - \frac{v_p^2 e^{2iE_k(t_1-t)}}{i(E_k - E_p - eV)}\Big)(f_p - f_k)$$

$$+ \Big(\frac{-v_p^2 e^{2iE_k(t-t_1)}}{i(E_k + E_p + eV)} - \frac{u_p^2 e^{2iE_k(t_1-t)}}{i(E_k + E_p - eV)}\Big)(1 - f_k - f_p)\Big]\Big\}\ . \tag{S38}$$

Finally, the integration over $t_1$ yields the static magnetization in the left lead,

$$\mathbf{M}_{L,\text{dc}} = \sum_{\mathbf{k},\mathbf{p}}\frac{\Delta_L^2}{4E_k^3}\sum_{\sigma,\sigma'}\boldsymbol{\sigma}_{\sigma\sigma'}\sum_{\sigma''}\Big(\mathcal{J}^*_{\mathbf{k}\sigma,\mathbf{p}\sigma''}\mathcal{J}_{\mathbf{k}\sigma',\mathbf{p}\sigma''} + \sigma\sigma'\mathcal{J}^*_{-\mathbf{k}-\sigma',\mathbf{p}\sigma''}\mathcal{J}_{-\mathbf{k}-\sigma,\mathbf{p}\sigma''}\Big)$$

$$\times\Big[\frac{f_p - f_k}{(E_k - E_p)^2 - (eV)^2}\Big(\frac{\xi_p}{E_p}(E_p - E_k) + eV\Big) + \frac{1 - f_k - f_p}{(E_k + E_p)^2 - (eV)^2}\Big(\frac{\xi_p}{E_p}(E_p + E_k) + eV\Big)\Big]\ . \tag{S39}$$

As seen, the discussion following Eq. (S23) for the oscillatory magnetization pertains also to the static one. However, the terms determining the spin structure are different. For instance, when the spin dependence of the tunneling amplitudes is due solely to a spin-orbit interaction in the weak link which obeys Eq. (S24), then

$$\sum_{\sigma,\sigma'}\boldsymbol{\sigma}_{\sigma\sigma'}\sum_{\sigma''}\Big(\mathcal{J}^*_{\mathbf{k}\sigma,\mathbf{p}\sigma''}\mathcal{J}_{\mathbf{k}\sigma',\mathbf{p}\sigma''} + \sigma\sigma'\mathcal{J}^*_{-\mathbf{k}-\sigma',\mathbf{p}\sigma''}\mathcal{J}_{-\mathbf{k}-\sigma,\mathbf{p}\sigma''}\Big) = \sum_{\sigma,\sigma'}\boldsymbol{\sigma}_{\sigma\sigma'}(1 + \sigma\sigma')\sum_{\sigma''}\mathcal{J}^*_{\mathbf{k}\sigma,\mathbf{p}\sigma''}\mathcal{J}_{\mathbf{k}\sigma',\mathbf{p}\sigma''}\ , \tag{S40}$$

and consequently has only the $\hat{\mathbf{z}}$ component, yielding

$$\sum_{\sigma,\sigma'}\boldsymbol{\sigma}_{\sigma\sigma'}(1 + \sigma\sigma')\sum_{\sigma''}\mathcal{J}^*_{\mathbf{k}\sigma,\mathbf{p}\sigma''}\mathcal{J}_{\mathbf{k}\sigma',\mathbf{p}\sigma''} = 2\hat{\mathbf{z}}(|\mathcal{J}_{\mathbf{k}\uparrow,\mathbf{p}\uparrow}|^2 - |\mathcal{J}_{\mathbf{k}\downarrow,\mathbf{p}\downarrow}|^2 + |\mathcal{J}_{\mathbf{k}\uparrow,\mathbf{p}\downarrow}|^2 - |\mathcal{J}_{\mathbf{k}\downarrow,\mathbf{p}\uparrow}|^2)\ . \tag{S41}$$

As is easily seen, for the two scenarios described above [see Eqs. (S26) and (S27)], the dc magnetization vanishes. However, applying a magnetic field at the interface changes the situation. In this case [S5]

$$\mathcal{J}_{-\mathbf{k}-\sigma,-\mathbf{p}-\sigma''}(B) = \mathcal{J}^*_{\mathbf{k}\sigma,\mathbf{p}\sigma''}(-B)\ , \tag{S42}$$

and consequently the spin factor becomes

$$\sum_{\sigma,\sigma'}\boldsymbol{\sigma}_{\sigma\sigma'}\sum_{\sigma''}\Big(\mathcal{J}^*_{\mathbf{k}\sigma,\mathbf{p}\sigma''}(B)\mathcal{J}_{\mathbf{k}\sigma',\mathbf{p}\sigma''}(B) + \sigma\sigma'\mathcal{J}^*_{-\mathbf{k}-\sigma',\mathbf{p}\sigma''}(B)\mathcal{J}_{-\mathbf{k}-\sigma,\mathbf{p}\sigma''}(B)\Big)$$

$$= \sum_{\sigma,\sigma'}\boldsymbol{\sigma}_{\sigma\sigma'}\sum_{\sigma''}\Big(\mathcal{J}^*_{\mathbf{k}\sigma,\mathbf{p}\sigma''}(B)\mathcal{J}_{\mathbf{k}\sigma',\mathbf{p}\sigma''}(B) + \sigma\sigma'\mathcal{J}_{\mathbf{k}\sigma',\mathbf{p}\sigma''}(-B)\mathcal{J}^*_{\mathbf{k}\sigma,\mathbf{p}\sigma''}(-B)\Big)\ . \tag{S43}$$

As an example, we consider the tunneling amplitude introduced in Ref. [S5] (see in particular the supplemental material). For a straight weak link along the $\hat{\mathbf{x}}$ direction, in which the electrons are subjected to a Rashba interaction $k k_{\rm so} \sigma_z/m^*$ ($m^*$ is the effective mass) and a Zeeman field $B_y$ (in energy units) along the $\hat{\mathbf{y}}$ direction, the tunneling amplitude, to linear order in $B_y$ takes the form

$$\boldsymbol{\mathcal{J}}(B_y) = \frac{im^*d}{2\sqrt{k_{\rm F}^2+k_{\rm so}^2}} e^{i\sqrt{k_{\rm F}^2+k_{\rm so}^2}d}\Big(-e^{ik_{\rm so}d}\Big[1+\sigma_z+\frac{m^*B_y\sigma_y}{k_{\rm so}(\sqrt{k_{\rm F}^2+k_{\rm so}^2}+k_{\rm so})}\Big]$$

$$+ e^{-ik_{\rm so}d}\Big[-1+\sigma_z+\frac{m^*B_y\sigma_y}{k_{\rm so}(\sqrt{k_{\rm F}^2+k_{\rm so}^2}-k_{\rm so})}\Big]\Big)$$

$$= \frac{im^*d}{2\sqrt{k_{\rm F}^2+k_{\rm so}^2}} e^{i\sqrt{k_{\rm F}^2+k_{\rm so}^2}d}\Big(-\cos(k_{\rm so}d)-i\sin(k_{\rm so}d)\sigma_z+\sigma_y\frac{m^*B_y}{k_{\rm F}^2}\Big[\cos(k_{\rm so}d)-i\frac{\sqrt{k_{\rm F}^2+k_{\rm so}^2}}{k_{\rm so}}\sin(k_{\rm so}d)\Big]\Big) . \quad (S44)$$

[Here we have used Eqs. (4-8) in the supplemental material of Ref. [S5] with $B_z=0$ and without the $B_y^2$ terms.] The prefactor is the bare tunneling amplitude $J_0$; hence,

$$\boldsymbol{\mathcal{J}}(B_y)/J_0 = -\cos(k_{\rm so}d)-i\sin(k_{\rm so}d)\sigma_z+\sigma_y\frac{m^*B_y}{k_{\rm F}^2}\Big[\cos(k_{\rm so}d)-i\frac{\sqrt{k_{\rm F}^2+k_{\rm so}^2}}{k_{\rm so}}\sin(k_{\rm so}d)\Big] , \quad (S45)$$

and (to linear order in $B_y$)

$$\frac{1}{|J_0|^2}[\boldsymbol{\mathcal{J}}(B_y)\boldsymbol{\mathcal{J}}^\dagger(B_y)]_{\sigma',\sigma} = \delta_{\sigma,\sigma'} - \frac{2m^*B_y}{k_{\rm F}^2}\cos(k_{\rm so}d)[\sin(k_{\rm so}d)\sigma_x+\cos(k_{\rm so}d)\sigma_y]_{\sigma',\sigma} . \quad (S46)$$

Obviously, only the off-diagonal elements can contribute. For those, the product $\sigma\sigma'$ in Eq. (S43) is negative, but since also $B_y$ changes sign we find that the spin factor is

$$\frac{8m^*B_y}{k_{\rm F}^2}\cos(k_{\rm so}d)[\sin(k_{\rm so}d)\hat{\mathbf{x}}+\cos(k_{\rm so}d)\hat{\mathbf{y}}] . \quad (S47)$$

The non-spin-dependent part of Eq. (S39) for $\mathbf{M}_{L,\rm dc}$ is (under the same assumptions as used above)

$$-\mu_B(eV)\sum_{\mathbf{k},\mathbf{p}}\frac{\Delta_L^2}{4E_k^3}\frac{1}{(E_k+E_p)^2} \approx -\mu_B\frac{eV}{\Delta}\frac{1}{16}\frac{R_Q}{R_n}$$

As we see from the example Eq. (S47) we should add a factor $4(\mu_B B_y)/E_F$ so that the estimate for the amplitude of the dc magnetization is

$$\mu_B\frac{eV}{\Delta}\frac{R_Q}{R_n}\frac{\mu_B B_y}{E_F}$$

which is very small.

---